\documentclass[pra,twocolumn,showpacs,preprintnumbers,amsmath,amssymb]{revtex4}
\usepackage{}

\usepackage{bm}% bold math
\usepackage{bbm}
\usepackage{amssymb}
\usepackage{amsfonts}
\usepackage{epsfig,graphicx}
\usepackage{amstext}
\usepackage{amsmath}
\usepackage{graphicx}
\usepackage{times}
\usepackage{txfonts}
\usepackage{dcolumn}
\usepackage{color}
\usepackage{dsfont}

\newcommand{\iden}{\mathds{1}}
\newcommand{\tr}{\mathrm{tr}}

\begin{document}
%%%%%%%%%%%%%%%%%%%%%%%%%%%%%%%%%%%%%%%%%%%%%%%%%%%%%%%%%%%%%%%%%%%%%%
%TCIDATA{OutputFilter=Latex.dll}
%TCIDATA{Version=5.00.0.2552}
%TCIDATA{<META NAME="SaveForMode" CONTENT="1">}
%TCIDATA{LastRevised=Wednesday, June 22, 2005 16:21:09}
%TCIDATA{<META NAME="GraphicsSave" CONTENT="32">}

\title{Maximal steered coherence and its conversion to entanglement in multiple bosonic reservoirs}

\author{Xiao-Xiao Xu}
\affiliation{School of Science, Xi'an University of Posts and Telecommunications, Xi'an 710121, China}

\author{Ming-Liang Hu}
\email{mingliang0301@163.com}
\affiliation{School of Science, Xi'an University of Posts and Telecommunications, Xi'an 710121, China}

\begin{abstract}
%%%%%%%%%%%%%%%%%%%%%%%%%%%%%%%%%%%%%%%%%%%%%%%%%%%%%%%%%%%%%%%%%%
The remote control of coherence is a crucial step for its
application in quantum computation. We investigate the maximal
steered coherence (MSC) and its conversion to entanglement in
multiple bosonic reservoirs. It is shown that the MSC decays with
time in the Markovian regime and behaves as damped oscillations in
the non-Markovian regime. The MSC can also be connected directly
to the strength of non-Markovianity, through which we show that it
can be noticeably enhanced by taking full advantage of the
non-Markovian effects. Besides, the MSC can be converted
completely to entanglement via optimal incoherent operation
applied to the steered qubit and an incoherent ancilla which is
immune of decoherence, and the generated entanglement is stronger
than the shared entanglement in prior. These findings suggest a
potential way for remotely generating and manipulating coherence
and entanglement in noisy environments.
\end{abstract}


\maketitle

\section{Introduction} \label{sec:1}
%%%%%%%%%%%%%%%%%%%%%%%%%%%%%%%%%%%%%%%%%%%%%%%%%%%%%%%%%%%%%%%%%%
As the fundamental characteristics differentiating a quantum
system from its classical counterpart, quantum coherence has
remained one of the research focuses of the quantum community for
over a century \cite{Ficek}. In the past few decades, quantum
coherence further found its application in the emerging fields of
quantum computation, quantum communication, and quantum metrology
\cite{Nielsen}. Recently, it is attracting growing interest once
again, motivated by the formulation of the resource theory of
coherence which sets the stage for analyzing quantitatively the
decoherence mechanism of the open systems \cite{Plenio,Hu,coher}.
Within this framework, a number of coherence measures have also
been introduced \cite{coher,mea1,mea2,mea3,mea4,mea5,mea6},
starting from which researchers further studied explicitly its
role in the tasks of state merging \cite{qsm}, deterministic
quantum computation with one qubit \cite{DQC1}, phase
discrimination \cite{mea4,metr}, subchannel discrimination
\cite{mea5}, and the Deutsch-Jozsa algorithm \cite{DJ}.

Although quantum coherence is defined for a single-partite system,
it is intimately related to different quantum correlations. Their
interrelations could be revealed by dividing a composite system
into different subsystems and then analyzing distribution of the
coherence among these subsystems \cite{mea1,mea6,coen3,coqd1,coqd2}.
From a practical point of view, it is also meaningful to explore
their interrelations from an operational perspective, e.g., by considering the
steered coherence of a bipartite system via local operations and
classical communication (LOCC) \cite{msc,Hux,asc1,asc2,asc3}.
Actually, the coherence distillation and dilution tasks have been
extensively studied based on different quantum operations
\cite{dist1,dist2,dist3,dist4,dist5,dist6}. For a bipartite system
$AB$, it has also been shown that one cannot create coherence on
$B$ via LOCC only when $\rho_{AB}$ is a quantum-incoherent state
\cite{rcc}. For the steered coherence averaged over all the
mutually unbiased bases, it has been shown that it may capture a
kind of quantum correlation stronger than entanglement
\cite{asc1,asc2}. Moreover, the maximal steered coherence (MSC)
can be linked to discord-like correlations, e.g., it vanishes
for the zero-discord states \cite{msc}. Here, by saying the
MSC, we mean the steered coherence on $B$ maximized over the
positive-operator-valued measure (POVM) on $A$.

Quantum coherence of a system $S$ could be converted to
entanglement via incoherent operations on $S$ and an ancilla
\cite{mea1}. Moreover, the amount of coherence in a state can be
enhanced by performing a unitary operation on it \cite{mc1,mc2,mc3}.
Thus for a bipartite system $AB$ shared between two participants,
one might realize the control of coherence on $B$ via LOCC and the
controlled creation of entanglement between $B$ and an incoherent
ancilla $C$ by utilizing the MSC (or the MSC further maximized
over the unitary transformations) on $B$. Besides, the ancillary
system can be initialized in a state that is immune of decoherence
in general, so it may be able to suppress the detrimental effect
of decoherence on entanglement.

In this paper, we investigate the MSC of two qubits $A$ and $B$
immersed in two groups of multiple bosonic reservoirs, aimed at
revealing the (non-)Markovian effects triggered by different
physical mechanisms on its behaviors and its conversion to
entanglement. We will show that the MSC decays exponentially with
time in the Markovian regime and behaves as damped oscillations of
time in the non-Markovian regime. Besides, the entanglement
converted from the MSC is stronger than the entanglement of
$AB$. This observation manifests efficiency of the active quantum
operations on remote control of coherence and entanglement in
noisy environments.

This paper is arranged as follows. In Section \ref{sec:2}, we recall
measures of the MSC, then in Section \ref{sec:3}, we give
solution of the multiple reservoirs and the non-Markovianity. In
Sections \ref{sec:4} and \ref{sec:5}, we present an analysis of the
decay behaviors of the MSC and a comparison of the entanglement
converted from the MSC with the shared entanglement of two
participants. Finally, Section \ref{sec:6} is devoted to a summary
of the main results.

\section{Maximal steered coherence} \label{sec:2}
%%%%%%%%%%%%%%%%%%%%%%%%%%%%%%%%%%%%%%%%%%%%%%%%%%%%%%%%%%%%%%%%%%
We consider the steering of coherence between two participants
Alice (\textit{A}) and Bob (\textit{B}) who share a two-qubit
state $\rho_{AB}$ in prior. First, Alice performs the POVM
measurements $M$ on qubit \textit{A}, after which the state of
qubit \textit{B} collapses to
%%%%%%%%%%%%%%%%%%%%%%%%%%%%%
\begin{equation}\label{eq2-1}
 \rho_{B|M}= \tr_A(M\otimes\iden\rho_{AB})/p_M,
\end{equation}
%%%%%%%%%%%%%%%%%%%%%%%%%%%%%
where $p_M=\tr(M\otimes\iden\rho_{AB})$ is the probability for
attaining $\rho_{B|M}$. To quantify the net amount of coherence
generated on $B$, Bob chooses the reference basis
$\{|\xi_i\rangle\}$ spanned by the eigenbasis of $\rho_B=
\tr_A \rho_{AB}$ in which there is no initial coherence on $B$.
Hence, the MSC attainable by Bob will be given by
\cite{msc}
%%%%%%%%%%%%%%%%%%%%%%%%%%%%%
\begin{equation}\label{eq2-2}
 C_\mu^{\mathrm{msc}}(\rho_{AB})= \inf_{\{|\xi_i\rangle\}}\left\{\max_{M\in\mathrm{POVM}}
                                     C_\mu^{\{|\xi_i\rangle\}}(\rho_{B|M})\right\},
\end{equation}
%%%%%%%%%%%%%%%%%%%%%%%%%%%%%
where the infimum over $\{|\xi_i\rangle\}$ is introduced as
$\rho_B$ may be degenerate, while $C_\mu^{\{|\xi_i\rangle\}}
(\rho_{B|M})$ represents a measure of coherence in $\rho_{B|M}$.
In this paper, we will consider two well-accepted measures of
coherence, that is, the $l_1$ norm of coherence and the relative
entropy of coherence given by \cite{coher}
%%%%%%%%%%%%%%%%%%%%%%%%%%%%%%
\begin{equation}\label{eq2-3}
\begin{aligned}
 & C_{l_1}^{\{|\xi_i\rangle\}}(\rho_{B|M})= \sum_{i\neq j} |\langle\xi_i|\rho_{B|M}|\xi_j\rangle|,\\
 & C_\mathrm{re}^{\{|\xi_i\rangle\}}(\rho_{B|M})= S[(\rho_{B|M})_{\mathrm{diag}}]-S(\rho_{B|M}),
\end{aligned}
\end{equation}
%%%%%%%%%%%%%%%%%%%%%%%%%%%%%
where $(\rho_{B|M})_{\mathrm{diag}}$ denotes the state obtained
by deleting all the off-diagonal elements of $\rho_{B|M}$, and
$S(\cdot)$ is the von Neumann entropy of the corresponding state.
Of course, the MSC can also be quantified by other coherence
measures, but their behaviors are qualitatively the same, thus we
will do not consider them one by one.

\section{Solution of the model} \label{sec:3}
\begin{figure}
\centering
\resizebox{0.44 \textwidth}{!}{%
\includegraphics{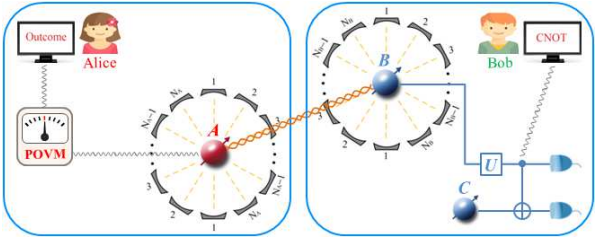}}
% If not, use\vspace{5cm}
% Give the correct figure height in cm
\caption{Schematic picture for steering Bob's coherence via LOCC in
multiple bosonic reservoirs (which could be realized, e.g., by two
sets of lossy cavities) and conversion of the MSC to entanglement
of \textit{BC} under the \textsc{cnot} operation. Here, the control
qubit \textit{B} is entangled with qubit \textit{A}, while the
target qubit \textit{C} is initialized in the ground state
$|0\rangle$ which is immune of the bosonic reservoirs.} \label{fig:1}
% Give a unique label
\end{figure}

We consider a central system consists of two noninteracting
two-level atoms (serve as the qubits) labeled as \textit{A} and
\textit{B}. When they are coupled independently to two sets of
multiple bosonic reservoirs as shown in Fig. \ref{fig:1}, the total
Hamiltonian is given by $\hat{H}=\hat{H}_A+\hat{H}_B$, where the
single ``qubit+reservoir" Hamiltonian $\hat{H}_S$ ($S=A$ or $B$)
reads (in units of $\hbar$)
%%%%%%%%%%%%%%%%%%%%%%%%%%%%%
\begin{equation}\label{eq3-1}
 \hat{H}_S= \frac{1}{2}\omega_0\sigma_z + \sum_{n=1}^{N_S}
            \sum_k \big[\omega_{n,k}b_{n,k}^\dag b_{n,k}
            +g_{n,k} (b_{n,k} \sigma_{+}+\mathrm{H.c.}) \big],
\end{equation}
%%%%%%%%%%%%%%%%%%%%%%%%%%%%%
where $\omega_0$ is the transition frequency between the ground
state $|0\rangle$ and the excited state $|1\rangle$ of the two qubits,
$\sigma_\pm=(\sigma_x\pm i\sigma_y)/2$ are the raising and
lowering operators, $\sigma_{x,y,z}$ are the three Pauli operators,
and $b_{n,k}$ ($b_{n,k}^\dag$) is the annihilation (creation)
operator of the $n$th reservoir's field mode $k$ with frequency
$\omega_{n,k}$, while its coupling strength to the qubit is
$g_{n,k}$. Moreover, $N_S$ is the number of reservoirs acting on
$S$, which may be implemented by $N_S$ pairs of lossy cavity
mirrors \cite{mirror,Maniscalco,manzx} as sketched in Fig.
\ref{fig:1}.

For the qubit $S$ ($S=A$ or $B$) being prepared initially in the
state $\rho_{S}(0)$ and there is no initial correlation between
$S$ and the reservoirs, its evolved state after tracing over the
$N_S$ reservoirs can be obtained as
\cite{Maniscalco,manzx}
%%%%%%%%%%%%%%%%%%%%%%%%%%%%%
\begin{equation}\label{eq3-2}
\rho_S(t)=\left(\begin{array}{cc}
 \rho_S^{11}(0)|p_S(t)|^2  & \rho_S^{10}(0)p_S (t) \vspace{0.5em}\\
 \rho_S^{01}(0)p_S^*(t)    & 1-\rho_S^{11}(0)|p_S(t)|^2
\end{array}\right),
\end{equation}
%%%%%%%%%%%%%%%%%%%%%%%%%%%%%
where $\rho_S^{ij}(0)=\langle i|\rho_S(0)|j\rangle$, and $p_S(t)$
is a time-dependent parameter determined by the spectra of these
reservoirs. We will consider the reservoirs with the Lorentzian spectrum
$J_n(\omega)$ for which $p_S(t)$ is analytically solvable. Here,
$J_n(\omega)= \gamma_n\lambda_n^2/\{[2\pi[(\omega-\omega_0)^2+\lambda_n^2]\}$,
where $\lambda_n$ denotes the spectral width of the $n$th reservoir
whose reciprocal determines its characteristic correlation time,
while the reciprocal of $\gamma_n$ determines the relaxation time of
the qubit \cite{Breuer}. In the following, we focus on the case that
all the reservoirs are the same (i.e., $\lambda_n\equiv\lambda$ and
$\gamma_n \equiv \gamma$, $\forall n$), then in the non-Markovian
regime, one has
%%%%%%%%%%%%%%%%%%%%%%%%%%%
\begin{equation}\label{eq3-3}
 p_S(t)= e^{-\frac{1}{2}\lambda t} \left(\cos\frac{d_S t}{2}+
         \frac{\lambda}{d_S}\sin\frac{d_S t}{2}\right),
\end{equation}
%%%%%%%%%%%%%%%%%%%%%%%%%%%
where $d_S= (|\lambda^2-2 N_S\gamma\lambda|)^{1/2}$. In the
Markovian regime, $p_S(t)$ has a similar form but with
$\cos(\cdot)$ and $\sin(\cdot)$ being replaced by $\cosh(\cdot)$
and $\sinh(\cdot)$, respectively. Here, the (non-)Markovianity
can be distinguished by the relative magnitudes of $\lambda$
and $\gamma$. For $\lambda> 2 N_S\gamma$, $p_S(t)$ decays
exponentially with time and the evolution is Markovian. For
$\lambda< 2 N_S\gamma$, $p_S(t)$ oscillates with time which is
an embodiment of the non-Markovian dynamics.

For given $\lambda$ and $\gamma$,
one can see from Eq. \eqref{eq3-3} that there exists a
critical value for the number $N_S$ of reservoirs after which the
non-Markovianity occurs. Such a critical value can be obtained as
$N_{S,\mathrm{cr}}= \lfloor\lambda/2\gamma\rfloor+1$ ($\lfloor x
\rfloor$ is the nearest integer not larger than $x$).
The non-Markovianity occurs when $N_S \geqslant N_{S,\mathrm{cr}}$
can be confirmed via the Breuer-Laine-Piilo measure of
non-Markovianity defined as $\mathcal{N}_\mathrm{BLP}=
\max_{\rho_{1,2}(0)} \int_{\varepsilon>0} \varepsilon
[t,\rho_{1,2}(0)]\mathrm{d}t$ \cite{BLP}, where $\varepsilon
[t,\rho_{1,2}(0)]$ is the time derivative of the trace distance
between $\rho_1(t)$ and $\rho_2(t)$ defined by $D[\rho_1(t),\rho_2(t)]
=\tr |\rho_1(t)-\rho_2(t)|/2$. As has been shown in Ref.
\cite{manzx}, for the optimal $\rho_{1,2}^\mathrm{op}(0)$ (i.e.,
the eigenstates of $\sigma_x$), one has $D[\rho_1(t),\rho_2(t)]
=|p_S(t)|$, which vanishes at the critical times $t_{z,l}=
2(l\pi-\vartheta_S)/d_S$ and reaches to its peak at $t_{p,l}=
2(l-1)\pi/d_S$, with $\vartheta_S= \arctan(d_S/\lambda)$ and
$l\in\mathbb{N}$. From these results one can obtain
$\mathcal{N}_\mathrm{BLP}=\sum_{l=1}^\infty e^{-l\lambda\pi/d_S}$.
It increases with the increase of $N_S$ when $N_S \geqslant
N_{S,\mathrm{cr}}$. Besides, one can also quantify the non-Markovianity
based on the backflow ratio of information, which we define it as
%%%%%%%%%%%%%%%%%%%%%%%%%%%
\begin{equation}\label{eq3-4}
 \mathcal{N}_\mathrm{BRI}= \frac{\int_{\varepsilon>0} \varepsilon[t,\rho_{1,2}^\mathrm{op}(0)]\mathrm{d}t}
                                {\int_{\varepsilon<0} \varepsilon[t,\rho_{1,2}^\mathrm{op}(0)]\mathrm{d}t},
\end{equation}
%%%%%%%%%%%%%%%%%%%%%%%%%%%
which is the ratio of the time-integration of $\varepsilon$ over all
the time intervals $[t_{z,l},t_{p,l+1}]$ in which $\varepsilon$ is
positive and that over all the time intervals $[t_{p,l},t_{z,l}]$
in which $\varepsilon$ is negative. Here, one may interpret the
term on the numerator (denominator) of Eq. \eqref{eq3-4} as the
backflow (outflow) of information. Different from
$\mathcal{N}_\mathrm{BLP}$ which contains a summation over all
$l\in \mathbb{N}$, one can show that in every time interval
$[t_{p,l},t_{p,l+1}]$ ($l\in\mathbb{N}$), the ratio of the
time-integration of the two terms in Eq. \eqref{eq3-4} is
completely the same. Therefore, one has the following exact result:
%%%%%%%%%%%%%%%%%%%%%%%%%%%
\begin{equation}\label{eq3-5}
 \mathcal{N}_\mathrm{BRI}= e^{-\lambda\pi/d_S},
\end{equation}
%%%%%%%%%%%%%%%%%%%%%%%%%%%
which is also larger than zero when $N_S\geqslant N_{S,\mathrm{cr}}$.
Besides, it also increases with an increase in $N_S$ when
$N_S\geqslant N_{S,\mathrm{cr}}$.

\section{MSC in multiple bosonic reservoirs} \label{sec:4}
%%%%%%%%%%%%%%%%%%%%%%%%%%%%%%%%%%%%%%%%%%%%%%%%%%%%%%%%%%%%%%%%%%
While the authors in Ref. \cite{manzx} considered the issue
of a single qubit transversally coupled to the multiple bosonic
reservoirs, it is also meaningful to generalize this theoretical
model to the two-qubit case for which one can further investigate
effects of the multiple bosonic reservoirs on controlling quantum
correlations.
As a matter of fact, for two qubits coupled independently to two
bosonic reservoirs with $N_{A}=N_{B}=1$, the dynamical behaviors
of entanglement \cite{Maniscalco,Bellomo,entan1,entan2},
discord-like correlations \cite{loren1,loren2,loren3}, and entropic
uncertainty relation \cite{eur}, have already been investigated,
and it is found that the non-Markovian effect triggered by
increasing the coupling strength is beneficial to them. Then it
is natural to ask whether the non-Markovian effect triggered by
increasing the number of reservoirs acting on each qubit is
beneficial for protecting coherence and entanglement of a state.

\begin{figure}
\centering
\resizebox{0.44 \textwidth}{!}{%
\includegraphics{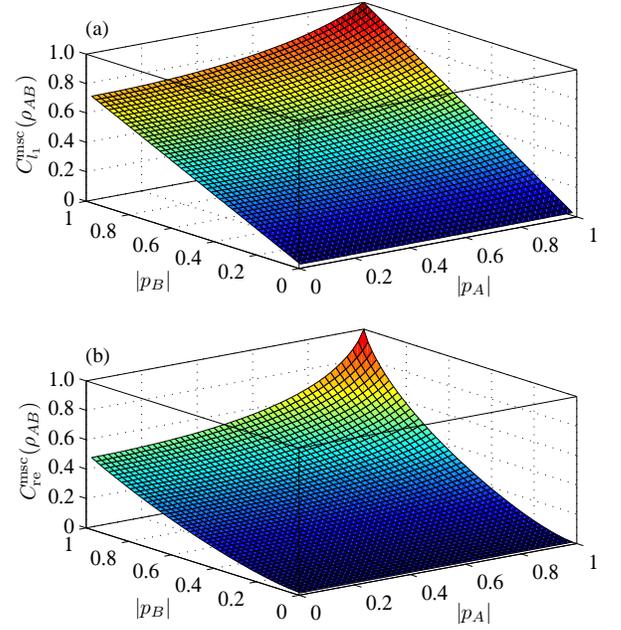}}
% If not, use\vspace{5cm}
% Give the correct figure height in cm
\caption{The MSC $C_\mu^\mathrm{msc}(\rho_{AB})$ ($\mu=l_1$ or
$\mathrm{re}$) versus $|p_A|$ and $|p_B|$ for the initial state
$|\Psi^+\rangle$ of qubits $A$ and $B$.} \label{fig:2}
% Give a unique label
\end{figure}

In this section, we explore behaviors of MSC for two qubits coupled
independently to two groups of multiple reservoirs as shown in Fig.
\ref{fig:1}. We will consider two slightly different cases: the case
of $N_A=N_B$ for which we call it symmetric reservoirs and the case
of $N_A\neq N_B$ for which we call it asymmetric reservoirs. We
consider the initial Bell-like state $|\Psi\rangle=\alpha|10\rangle
+\beta|01\rangle$ ($|\alpha|^2+|\beta|^2=1$).
Note that although the two qubits might be initially quantum
correlated depending on the parameters $\alpha$ and $\beta$, there
is no direct interaction neither between the two qubits nor between
the two multiple bosonic reservoirs.
Hence, one can derive the evolving state of the two qubits
based on the method given in Ref. \cite{Bellomo}. To be explicit,
we write the elements of $\rho_S(t)$ of Eq. \eqref{eq3-2} as
$\rho_S^{ii'}(t)=\sum_{ll'} S_{ii'}^{ll'}(t) \rho_S^{ll'}(0)$
($i,i',l,l'\in\{0,1\}$), from which one can obtain the nonzero
$S_{ii'}^{ll'}(t)$ as
%%%%%%%%%%%%%%%%%%%%%%%%%%%
\begin{equation}\label{eq4-0}
 \begin{aligned}
  & S_{11}^{11}(t)=|p_S|^2,~ S_{10}^{10}(t)=p_S,~ S_{01}^{01}(t)=p_S^*, \\
  & S_{00}^{00}(t)=1,~ S_{00}^{11}(t)=1-|p_S|^2,
 \end{aligned}
\end{equation}
%%%%%%%%%%%%%%%%%%%%%%%%%%%
where $S=A$ or $B$. By substituting these into Eq. (5) of Ref.
\cite{Bellomo} and using the fact that $p_S\in\mathbb{R}$, one can obtain
%%%%%%%%%%%%%%%%%%%%%%%%%%%
\begin{equation}\label{eq4-1}
\rho_{AB}(t)= \left(\begin{array}{cccc}
 0  & 0                      & 0                      & 0 \vspace{0.5em}\\
 0  & |\alpha p_A|^2         & \alpha\beta^* p_A p_B  & 0 \vspace{0.5em}\\
 0  & \alpha^*\beta p_A p_B  & |\beta p_B|^2          & 0  \vspace{0.5em}\\
 0  & 0                      & 0                      & 1-|\alpha p_A|^2-|\beta p_B|^2
\end{array}\right),
\end{equation}
%%%%%%%%%%%%%%%%%%%%%%%%%%%
and we will focus on the case of $\alpha\beta\neq 0$ unless
specifically stated based on the consideration that Alice cannot
steer Bob's coherence at all when $\alpha\beta= 0$.

For the two-qubit states, as has been explained in Ref. \cite{msc},
one only needs to take the maximization over the set of projective
measurements $M=(\iden+ \vec{m} \cdot \bm{\sigma})/2$, where
$\bm{\sigma}= (\sigma_x,\sigma_y,\sigma_z)$ and
$\vec{m}= (\sin\theta\cos\phi,\sin\theta\sin\phi,\cos\theta)$,
with $\theta$ and $\phi$ being the polar and azimuth angles,
respectively. Then the postmeasurement state of qubit $B$ can be
obtained as
%%%%%%%%%%%%%%%%%%%%%%%%%%%
\begin{equation}\label{eq4-2}
\rho_{B|M} =\left(\begin{array}{cc}
 \frac{|\beta p_B|^2 \sin^2(\theta/2)} {p_M} &
 \frac{e^{-i\phi}\alpha^* \beta p_A p_B\sin\theta} {2p_M} \vspace{0.5em}\\
 \frac{e^{i\phi}\alpha \beta^* p_A p_B\sin\theta} {2p_M} &
 1-\frac{|\beta p_B|^2\sin^2(\theta/2)} {p_M}
\end{array}\right),
\end{equation}
%%%%%%%%%%%%%%%%%%%%%%%%%%%
where $p_M=[1+ (2|\alpha p_A|^2-1)\cos\theta]/2$, so the $l_1$
norm of MSC remains zero for $\alpha\beta=0$, irrespective of
$\theta$ and $\phi$. Otherwise, the optimal polar angle is
$\theta_0= \arccos(1-2|\alpha p_A|^2)$, while the azimuth angle
can take any value. As a result, one has
%%%%%%%%%%%%%%%%%%%%%%%%%%%
\begin{equation}\label{eq4-3}
  C_{l_1}^\mathrm{msc}(\rho_{AB})= \left\{
  \begin{aligned}
   & \frac{|\beta p_B|}{\sqrt{1-|\alpha p_A|^2}} &\text{if}~\alpha\beta\neq 0,\\
   & 0 &\text{if}~\alpha\beta=0,
  \end{aligned} \right.
\end{equation}
%%%%%%%%%%%%%%%%%%%%%%%%%%%
from which one can note that it is monotonic increasing functions
of both $|p_A|$ and $|p_B|$. As has been shown in Section \ref{sec:3},
the decrease of $|p_S|$ ($S=A$ or $B$) signifies an outflow of
information to the reservoirs, while the increase of $|p_S|$
signifies a backflow of information to the system. Hence, for the
initial state $|\Psi\rangle$ with $\alpha\beta \neq 0$, the backflow
of information to either one of the multiple reservoirs is always
beneficial for protecting the $l_1$ norm of MSC. But the decay
rates with respect to $|p_A|$ and $|p_B|$ are somewhat different.
More specifically, for the symmetric multiple reservoirs, one can obtain
from Eq. \eqref{eq4-3} that the decay rate of the $l_1$ norm of MSC
with respect to $|p_A|$ ($p_B=p_A$) increases with the increase of
$|p_A|$. For the asymmetric multiple reservoirs, it decays linearly
with the decrease of $|p_B|$ and the decay rate is independent of
$|p_B|$, while for $p_B\neq 0$, it shows a parabolic decrease with
the decrease of $|p_A|$, but the decay rate is not a constant
function of $|p_A|$, see Fig. \ref{fig:2}(a).

As for the relative entropy of MSC, it is also independent of the
azimuth angle $\phi$, but it is difficult to obtain analytically
the optimal polar angle $\theta_0$. So we have to optimize it
numerically. Here, $ C_\mathrm{re}^\mathrm{msc}(\rho_{AB})$ can
be written as
%%%%%%%%%%%%%%%%%%%%%%%%%%%
\begin{equation} \label{eq4-4}
 C_\mathrm{re}^\mathrm{msc}(\rho_{AB})= \max_{\{\theta\}} \big\{S[(\rho_{B|M})_\mathrm{diag}]
                                        -S\left(\rho_{B|M}\right)\big\},
\end{equation}
%%%%%%%%%%%%%%%%%%%%%%%%%%%
and the corresponding numerical result is shown in Fig. \ref{fig:2}(b),
from which one can see that its dependence on $|p_A|$ and $|p_B|$ is
similar to that of the $l_1$ norm of MSC. The only difference is that
$ C_\mathrm{re}^\mathrm{msc}(\rho_{AB})$ does not decrease linearly
with the decreasing value of $|p_B|$.

\begin{figure}
\centering
\resizebox{0.44 \textwidth}{!}{%
\includegraphics{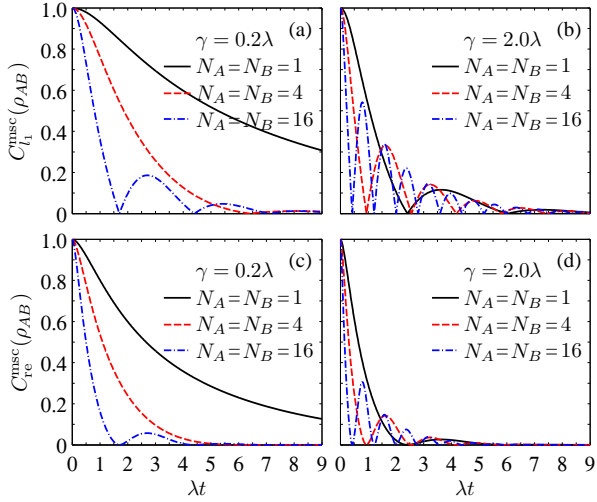}}
% If not, use\vspace{5cm}
% Give the correct figure height in cm
\caption{The MSC $C_\mu^\mathrm{msc}(\rho_{AB})$ ($\mu=l_1$ or
$\mathrm{re}$) versus $\lambda t$ for the initial state
$|\Psi^+\rangle$ with different $\gamma$, $N_A$, and $N_B$ in
the symmetric Lorentzian reservoirs. } \label{fig:3}
% Give a unique label
\end{figure}

In the following, we investigate the time dependence of the MSC. First,
we consider the case of the symmetric Lorentzian reservoirs. For the
initial state $|\Psi\rangle$ with $\alpha=\beta=1/\sqrt{2}$ (we
denote by it $|\Psi^+\rangle$ for conciseness of later presentation),
we show in Fig. \ref{fig:3} $C_\mu^\mathrm{msc}(\rho_{AB})$
($\mu=l_1$ or $\mathrm{re}$) versus $\lambda t$ with different
$\gamma$. For the given parameters in this figure, $N_{S,\mathrm{cr}}=3$
for $\gamma=0.2\lambda$ and $N_{S,\mathrm{cr}}=1$ for $\gamma=2.0\lambda$.
As expected, the MSC decreases monotonically with the evolving time
in the Markovian regime, see the solid lines in Fig. \ref{fig:3}(a)
and (c). When the non-Markovian effects are triggered, either by
increasing the coupling strength $\gamma$ or by increasing the number
of reservoirs, both the $l_1$ norm and relative entropy of MSCs
behave as damped oscillations (with the period $T=2\pi/d_S$) with
the time $t$ evolves and undergo sudden death at the critical times
$t_{z,l}$ ($l\in \mathbb{N}$). Physically, the decay behaviors are
caused by the exponential term in Eq. \eqref{eq3-3}, while the
oscillations are due to the sine and cosine terms. Moreover, at the
critical times $t_{p,l}$ ($l\in \mathbb{N}$), one has
%%%%%%%%%%%%%%%%%%%%%%%%%%%
\begin{equation}\label{eq4-5}
  C_{l_1}^\mathrm{msc}(\rho_{AB})\big|_{t=t_{p,l}}= \left\{
  \begin{aligned}
   & \frac{|\beta| \mathcal{N}_\mathrm{BRI}^{l-1}}{\sqrt{1-|\alpha|^2\mathcal{N}_\mathrm{BRI}^{2(l-1)}}} &\text{if}~\alpha\beta\neq 0,\\
   & 0 &\text{if}~\alpha\beta=0,
  \end{aligned} \right.
\end{equation}
%%%%%%%%%%%%%%%%%%%%%%%%%%%
which shows that the peak values of the MSC in the non-Markovian
regime can always be enhanced by increasing the non-Markovianity of the
multiple reservoirs.

\begin{figure}
\centering
\resizebox{0.44 \textwidth}{!}{%
\includegraphics{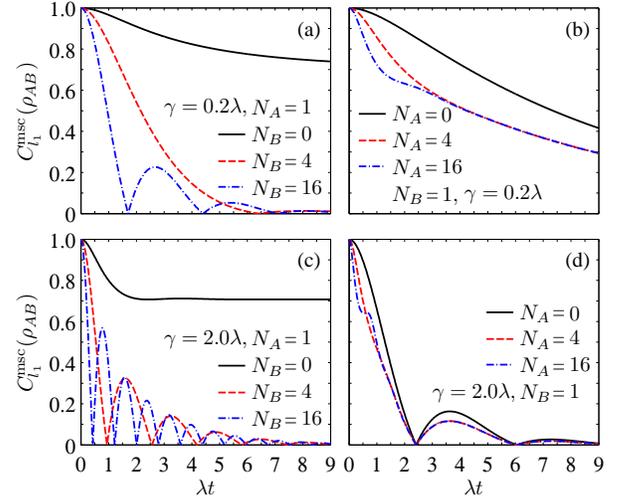}}
% If not, use\vspace{5cm}
% Give the correct figure height in cm
\caption{The MSC $C_{l_1}^\mathrm{msc}(\rho_{AB})$ versus $\lambda t$
for the initial state $|\Psi^+\rangle$ with different $\gamma$, $N_A$,
and $N_B$ in the asymmetric Lorentzian reservoirs.} \label{fig:4}
% Give a unique label
\end{figure}

Next, we consider the time dependence of the MSC in the asymmetric
Lorentzian reservoirs. In Fig. \ref{fig:4}, we show
$C_{l_1}^\mathrm{msc}(\rho_{AB})$ versus $\lambda t$ for
$|\Psi^+\rangle$ with different $\gamma$. When both $N_A$ and
$N_B$ are smaller than $N_{S,\mathrm{cr}}$, $|p_S(t)|$ shows a
Markovian exponential decay, hence as expected, the MSC also
decays exponentially with the time evolves, see the solid black
lines displayed in Fig. \ref{fig:4}. When one set of the multiple
reservoirs is in the non-Markovian regime, it can be found from
Fig. \ref{fig:4}(a) and (b) that for $N_A<N_B$, there are still
revives of the MSC, while for $N_A>N_B$ the MSC decays monotonically.
Such a difference is rooted in the asymmetric property of the MSC
and could be explained from Eq. \eqref{eq4-3}, as the enhancement
of $|p_A|$ may do not suffice to compensate the loss of $|p_B|$ for
the system parameters in Fig. \ref{fig:4}(b). Of course, one may
observe a very weak revival of the MSC by increasing slightly $N_B$
(e.g., $N_B=2$). When the two reservoirs acting on the qubits
\textit{A} and \textit{B} are already in the non-Markovian regime
for $N_{A,B}=1$, as can be found from Fig. \ref{fig:4}(c) and (d),
the MSC also behaves as damped oscillations with the time evolves.
But now the period and the peak values of the oscillations will be
determined by both $p_A$ and $p_B$.

For the relative entropy of MSC, its behaviors in the asymmetric
Lorentzian reservoirs are structurally similar to that of the
$l_1$ norm of MSC, so we do not present the corresponding plots
here.

As coherence itself is a precious resource for quantum computing
and the related quantum tasks, it is vital to seek flexible
methods to protect coherence. The previous studies showed that the
decoherence effects could be suppressed by taking full advantage
of the initial system-bath correlation \cite{Zhangyj}, the
correlations between consecutive actions of a noisy channel
\cite{deco,cpb}, and the quantum-jump-based feedback
\cite{feedback}. Besides, the coherence might also be frozen under
specific conditions \cite{fro1,fro2,fro3}. Different from these
methods for local controlling coherence, our results presented
above suggest a flexible way for remotely controlling coherence
in the noisy environments. In particular, such a steered
coherence may be stronger than that of the single-qubit
coherence under the same reservoirs. For example, if qubit $B$ is
uncorrelated with qubit $A$ (so Alice cannot steer its coherence)
and is initialized in the state $\alpha|1\rangle+\beta|0\rangle$,
the $l_1$ norm of coherence for the time-evolved state of $B$
will be given by $C_{l_1}[\rho_B(t)]=2|\alpha\beta^* p_B|$. It is
always weaker than the MSC of Eq. \eqref{eq4-3} when qubit $A$
is isolated from the reservoirs (i.e. $p_A=1$).  Even when qubit
$A$ is immersed in the bosonic reservoirs, $C_{l_1}[\rho_B(t)]$ is
still weaker than the MSC in the  parameter region of
$|\alpha|^2< [1-(1-p_A^2)^{1/2}]/2p_A^2$.

\section{Converting MSC to entanglement} \label{sec:5}
%%%%%%%%%%%%%%%%%%%%%%%%%%%%%%%%%%%%%%%%%%%%%%%%%%%%%%%%%%%%%%%%%%
Quantum coherence and entanglement are expensive resources
for quantum communication such as quantum cryptography \cite{new1},
motivated by which great efforts have been devoted to generating
long-lived coherence and entanglement in various physical systems,
e.g., the optomechanical system \cite{new2}, ultracold atomic
ensembles \cite{new3,new4}, and cold ion \cite{new5}.
Moreover, the coherence measures defined within the resource
theoretic framework could be measured experimentally with elaborately
designed techniques \cite{new6,new7} and are intimately
related to quantum entanglement \cite{Plenio,Hu,new8}.
In particular, as was shown in Ref. \cite{mea1}, by performing
the \textsc{cnot} operation $\Lambda_\textsc{cnot}$ on the qubit $B$
which is coherent and an ancillary qubit $C$ initialized in the
incoherent state $|0\rangle$, with $B$ ($C$) being the control
(target) qubit, these two qubits will become entangled. If one uses
concurrence \cite{EoF1,EoF2} as a measure of entanglement, then
$C\big(\Lambda_\textsc{cnot} [\rho_B\otimes |0\rangle\langle 0|]\big)
= C_{l_1}(\rho_B)$ \cite{mea1}, i.e., all the coherence in $\rho_B$
is converted to entanglement of $BC$ via the incoherent operation
$\Lambda_\textsc{cnot}$.

The above finding shows a way to control the entanglement of Bob's
qubits. For the scenario we considered in Fig. \ref{fig:1}, the
ancillary qubit $C$ is initialized in the ground state $|0\rangle$
and will be immune of the reservoirs. Hence, the generated entanglement
in $BC$ equals to the $l_1$ norm of coherence in $\rho_{B|M}$. For the
optimal state $\rho_{B|M,\mathrm{op}}$ (i.e., the state $\rho_{B|M}$
with $\theta= \theta_0$), one has
%%%%%%%%%%%%%%%%%%%%%%%%%%%
\begin{equation} \label{eq5-1}
 C\big(\Lambda_\textsc{cnot}[\rho_{B|M,\mathrm{op}}\otimes |0\rangle\langle 0|]\big)= C_{l_1}^\mathrm{msc}(\rho_{AB}),
\end{equation}
%%%%%%%%%%%%%%%%%%%%%%%%%%%
that is, the amount of created entanglement in $BC$ (measured
by concurrence) equals exactly to the $l_1$ norm of MSC for
$\rho_{AB}$. This gives an operational interpretation to the
MSC.

From Eq. \eqref{eq4-1} one can obtain the concurrence of
$\rho_{AB}$ as $C(\rho_{AB})=2|\alpha\beta^* p_A p_B|$. As a
result,
%%%%%%%%%%%%%%%%%%%%%%%%%%%
\begin{equation} \label{eq5-2}
 \frac{C(\rho_{AB})}{C\big(\Lambda_\textsc{cnot}[\rho_{B|M,\mathrm{op}}\otimes |0\rangle\langle 0|]\big)}
   = 2|\alpha p_A| \sqrt{1-|\alpha p_A|^2}
   \leqslant 1,
\end{equation}
%%%%%%%%%%%%%%%%%%%%%%%%%%%
thus the generated entanglement in $BC$ is always stronger than the
entanglement of $\rho_{AB}(t)$ whenever the reservoirs are present
(i.e., $|p_{A,B}|<1$). Even for the ideal case (i.e., $p_{A,B}=1$),
the equality holds only when $\rho_{AB}(t)$ belongs to one of the
Bell states. In Ref. \cite{new9}, a scheme for recovering
entanglement via local operations is proposed. Our scheme is
different from that in Ref. \cite{new9} as it suggests an efficient
way for remotely instead of locally manipulating entanglement in
noisy environments.

One might also concern the amount of generated entanglement in $BC$ without
Alice's steering. But for such a situation, one has $C(\Lambda_\textsc{cnot}
[\rho_B(t)\otimes |0\rangle\langle 0|])= 0$ for $\rho_{AB}(t)$
of Eq. \eqref{eq4-1}. This consolidates the fact that with the
help of the prior shared entanglement, Alice can control
efficiently the entanglement of Bob's qubits.

Of course, the enhancement of the generated entanglement in $BC$
compared to $C(\rho_{AB})$ is at the expense of reducing the
success probability. In fact, the optimal success probability
$p_{M,\mathrm{op}}$ (i.e., the probability of attaining
$\rho_{B|M,\mathrm{op}}$) is given by
%%%%%%%%%%%%%%%%%%%%%%%%%%%
\begin{equation} \label{eq5-3}
 p_{M,\mathrm{op}}= 2|\alpha p_A|^2(1-|\alpha p_A|^2),
\end{equation}
%%%%%%%%%%%%%%%%%%%%%%%%%%%
then one can see that the probability cannot exceed $50\%$.
To maximize the generated entanglement in $BC$, one can
choose the initial state $|\Psi^+\rangle$ for which
$|\alpha p_A|^2 \leqslant 1/2$. As a consequence,
$p_{M,\mathrm{op}}$ decreases with the decrease of $|p_A|$.
For the case of qubit \textit{A} being isolated perfectly
from the reservoirs, one has $p_{M,\mathrm{op}}=50\%$.

Note that the coherence of a state could be enhanced by performing
a unitary operation $U$ on it. For the single-qubit state $\rho$,
the maximal coherence under unitary operations is given by
$C_{l_1,\mathrm{op}}(\rho)=|\bm{x}|$, where $\bm{x}=(x_1,x_2,x_3)$,
with $x_i=\tr(\rho \sigma_i)$ and the optimal $U_\mathrm{op}$ can
be constructed by the eigenbasis of $\rho$ \cite{mc1,mc2}. For
$\rho_{B|M}$ of Eq. \eqref{eq4-2}, if Bob applies $U_\mathrm{op}$
to the qubit \textit{B} after Alice's optimal measurements (i.e.,
$\theta=\theta_0$), the optimized $l_1$ norm of MSC on $B$ will be
given by
%%%%%%%%%%%%%%%%%%%%%%%%%%%
\begin{equation} \label{eq5-4}
 C_{l_1,\mathrm{op}}^\mathrm{msc}(\rho_{B|M})= \sqrt{1-\frac{|\beta p_B|^2\big(1-|\alpha p_A|^2-|\beta p_B|^2\big)}{\big(1-|\alpha p_A|^2\big)^2}}.
\end{equation}
%%%%%%%%%%%%%%%%%%%%%%%%%%%
Such an optimized $l_1$ norm of MSC is stronger than that
of Eq. \eqref{eq4-3}, hence can be used to create stronger
entanglement in $BC$ than that given in Eq. \eqref{eq5-1}.
One can also note from the above equation that
$C_{l_1,\mathrm{op}}^\mathrm{msc}(\rho_{B|M})=1$ when $\beta=0$.
This is because for such a special case, $\rho_B(t)\equiv
|0\rangle\langle 0|$ and $U_\mathrm{op}$ transforms it to
the maximally coherent state \cite{Plenio}. But this lost
the original idea of steering as Alice cannot steer Bob's
coherence if $\alpha\beta=0$.

Finally, we would like to mention that when considering the
initial state $|\Phi\rangle=\alpha|11\rangle+ \beta|00\rangle$,
$\rho_{B|M}$ can be obtained in a similar way, from which one
can obtain that the $l_1$ norm of MSC is completely the same to
that for the initial state $|\Psi\rangle$. For the relative
entropy of MSC, although their strengths are somewhat different
for the initial states $|\Phi\rangle$ and $|\Psi\rangle$, their
behaviors are qualitatively the same. As a consequence, the
findings we presented above also apply to the case of two qubits
initialized in the Bell-like state $|\Phi\rangle$.

\section{Summary and discussion} \label{sec:6}
%%%%%%%%%%%%%%%%%%%%%%%%%%%%%%%%%%%%%%%%%%%%%%%%%%%%%%%%%%%%%%%%%%
To summarize, we have investigated the control of the MSC for two
qubits coupled independently to two groups of multiple bosonic
reservoirs. For the two qubits being initialized in the Bell-like
states, we showed that when the number of reservoirs acting on
each qubit is smaller than a critical value, the MSC shows an
exponential decay with time. But when it exceeds this critical
value, the MSC turns to behave as damped oscillations, where
the peak values of these oscillations can always be enhanced by
increasing the non-Markovianity of the reservoirs. Besides, apart
from the case of qubit \textit{B} being isolated perfectly from
the reservoirs, the MSC will decay to zero in the infinite-time
limit. Moreover, we have further shown that by performing the
\textsc{cnot} operation to the qubit \textit{B} and an ancillary
qubit $C$ initialized in the ground state which is immune of the
reservoirs, the MSC will be completely converted to entanglement
in $BC$, and such an entanglement generated with the help of LOCC
is stronger than or equals to the entanglement shared between
Alice and Bob. We hope these observations may shed some light on
revealing the interplay between the unavoidable decoherence of a
system and efficiency of the active quantum operations on
protecting coherence. From a practical point of view, it may
also provide an alternate for remotely generating and
manipulating coherence and entanglement in noisy environments.

Although we considered only the bosonic reservoirs with Lorentzian
spectrum, the investigations in this work can be generalized
immediately to other kinds of bosonic reservoirs, e.g., the sub-Ohmic,
Ohmic, and super-Ohmic reservoirs \cite{ohmic}, under which the
postmeasurement state of qubit $B$ has completely the same form
with that given in Eq. \eqref{eq4-2}, thus the dependence of MSC
on $|p_S|$ ($S=A$ or $B$) will also be the same. Of course, the
explicit time dependence of the MSC may be different due to the
different time dependencies of $p_S$ determined by the reservoir
spectral density.

%%%%%%%%%%%%%%%%%%%%%%%%%%%%%%%%%%%%%%%%%%%%%%%%%%%%%%%%%%%%%%%%%%
Finally, we remark that it is also of great interest to
further consider the finite temperature reservoirs. Although it is
hard to obtain the two-qubit density operator for this case due to
its complexity, one could give a heuristic analysis by considering
the Markovian case with $N_{A,B}=1$ for which $\rho_S(t)$ can be
obtained by solving the master equation in the Lindblad form, and
$\rho_{AB}(t)$ can be obtained in a similar way to that of Eq.
\eqref{eq4-1} (see Refs. \cite{finiteT1,finiteT2} for more detail).
Then for the initial Bell-like states, one can get a MSC behavior
qualitatively similar to what we have seen in Sections \ref{sec:4}
and \ref{sec:5}. The only difference is that the MSC is monotonically
decreased with an increase in the reservoir temperature. A general
study on the details of the finite temperature effects for the
non-Markovian multiple reservoirs is still needed.
Besides, it is also worthy to consider the case of two
qubits immersed in a common reservoir \cite{Breuer,loren3}, for
which there will be reservoir-mediated interaction between them.
The combined and intertwined effects of this indirect interaction
and non-Markovianity may induce more rich dynamics of the MSC than
that for two independent reservoirs, the details of which will be
considered elsewhere.

\section*{ACKNOWLEDGMENTS}
This work was supported by the National Natural Science Foundation
of China (Grant No. 11675129).

%%%%%%%%%%%%%%%%%%%%%%%%%%%%%%%%%%%%%%%%%%%%%%%%%%%%%%%%%%%%%%%%%%%%%%%
\newcommand{\PRL}{\emph{Phys. Rev. Lett.} }
\newcommand{\RMP}{\emph{Rev. Mod. Phys.} }
\newcommand{\PRA}{\emph{Phys. Rev. A} }
\newcommand{\PRB}{\emph{Phys. Rev. B} }
\newcommand{\PRD}{\emph{Phys. Rev. D} }
\newcommand{\PRE}{\emph{Phys. Rev. E} }
\newcommand{\PRX}{\emph{Phys. Rev. X} }
\newcommand{\APL}{\emph{Appl. Phys. Lett.} }
\newcommand{\NJP}{\emph{New J. Phys.} }
\newcommand{\JPA}{\emph{J. Phys. A} }
\newcommand{\JPB}{\emph{J. Phys. B} }
\newcommand{\OC}{\emph{Opt. Commun.} }
\newcommand{\PLA}{\emph{Phys. Lett. A} }
\newcommand{\EPJD}{\emph{Eur. Phys. J. D} }
\newcommand{\NP}{\emph{Nat. Phys.} }
\newcommand{\NPo}{\emph{Nat. Photonics} }
\newcommand{\NC}{\emph{Nat. Commun.} }
\newcommand{\EPL}{\emph{Europhys. Lett.} }
\newcommand{\AoP}{\emph{Ann. Phys. (N.Y.)} }
\newcommand{\QIC}{\emph{Quantum Inf. Comput.} }
\newcommand{\QIP}{\emph{Quantum Inf. Process.} }
\newcommand{\CPB}{\emph{Chin. Phys. B} }
\newcommand{\IJTP}{\emph{Int. J. Theor. Phys.} }
\newcommand{\IJQI}{\emph{Int. J. Quantum Inf.} }
\newcommand{\IJMPB}{\emph{Int. J. Mod. Phys. B} }
\newcommand{\PR}{\emph{Phys. Rep.} }
\newcommand{\SR}{\emph{Sci. Rep.} }
\newcommand{\LPL}{\emph{Laser Phys. Lett.} }
\newcommand{\SCG}{\emph{Sci. China Ser. G} }
\newcommand{\JSP}{\emph{J. Statis. Phys.} }
\newcommand{\SCPMA}{\emph{Sci. China-Phys. Mech. Astron.} }
\newcommand{\AQT}{\emph{Adv. Quantum Technol.} }

%BibTeX users please use
%\bibliographystyle{}
%\bibliography{}

\begin{thebibliography}{50}
%Format for Journal Reference
% 01-10
\bibitem{Ficek} Z. Ficek, S. Swain, \textit{Quantum Interference and Coherence: Theory and Experiments}, Springer Series in Optical Sciences, Springer, Berlin {\bf 2005}.
\bibitem{Nielsen} M. A. Nielsen, I. L. Chuang, \textit{ Quantum Computation and Quantum Information}, Cambridge University Press, Cambridge {\bf 2010}.
\bibitem{Plenio} A. Streltsov, G. Adesso, M. B. Plenio, \RMP {\bf 2017}, \emph{89}, 041003.
\bibitem{Hu} M. L. Hu, X. Hu, J. C. Wang, Y. Peng, Y. R. Zhang, H. Fan, \PR {\bf 2018}, \emph{762--764}, 1.
\bibitem{coher} T. Baumgratz, M. Cramer, M. B. Plenio, \PRL {\bf 2014}, \emph{113}, 140401.
\bibitem{mea1} A. Streltsov, U. Singh, H. S. Dhar, M. N. Bera, G. Adesso, \PRL {\bf 2015}, \emph{115}, 020403.
\bibitem{mea2} X. Yuan, H. Zhou, Z. Cao, X. Ma, \PRA {\bf 2015}, \emph{92}, 022124.
\bibitem{mea3} C. Napoli, T. R. Bromley, M. Cianciaruso, M. Piani, N. Johnston, G. Adesso, \PRL {\bf 2016}, \emph{116}, 150502. % metrology
\bibitem{mea4} K. Bu, U. Singh, S. M. Fei, A. K. Pati, J. Wu, \PRL {\bf 2017}, \emph{119}, 150405. % metrology
\bibitem{mea5} X. Qi, T. Gao, F. Yan, \JPA {\bf 2017}, \emph{50}, 285301.

% 11-20
\bibitem{mea6} K. Bu, N. Anand, U. Singh, \PRA {\bf 2018}, \emph{97}, 032342.
\bibitem{qsm} A. Streltsov, E. Chitambar, S. Rana, M. N. Bera, A. Winter, M. Lewenstein, \PRL {\bf 2016}, \emph{116}, 240405.
\bibitem{DQC1} J. Ma, B. Yadin, D. Girolami, V. Vedral, M. Gu, \PRL {\bf 2016}, \emph{116}, 160407.
\bibitem{metr} C. S. Yu, \PRA {\bf 2017}, \emph{95}, 042337. % metrology
\bibitem{DJ} M. Hillery, \PRA {\bf 2016}, \emph{93}, 012111.
\bibitem{coen3} K. C. Tan, H. Kwon, C. Y. Park, H. Jeong, \PRA {\bf 2016}, \emph{94}, 022329.
\bibitem{coqd1} Y. Yao, X. Xiao, L. Ge, C. P. Sun, \PRA {\bf 2015}, \emph{92}, 022112.
\bibitem{coqd2} M. L. Hu, H. Fan, \PRA {\bf 2017}, \emph{95}, 052106.
\bibitem{msc} X. Hu, A. Milne, B. Zhang, H. Fan, \SR {\bf 2015}, \emph{6}, 19365.
\bibitem{Hux} X. Hu, H. Fan, \SR {\bf 2016}, \emph{6}, 34380.

% 21-30
\bibitem{asc1} D. Mondal, T. Pramanik, A. K. Pati, \PRA {\bf 2017}, \emph{95}, 010301(R).
\bibitem{asc2} M. L. Hu, H. Fan, \PRA {\bf 2018}, \emph{98}, 022312.
\bibitem{asc3} M. L. Hu, X. M. Wang, H. Fan, \PRA {\bf 2018}, \emph{98}, 032317.
\bibitem{dist1} E. Chitambar, A. Streltsov, S. Rana, M. N. Bera, G. Adesso, M. Lewenstein, \PRL {\bf 2016}, \emph{116}, 070402.
\bibitem{dist2} A. Winter, D. Yang, \PRL {\bf 2016}, \emph{116}, 120404.
\bibitem{dist3} E. Chitambar, M.-H. Hsieh, \PRL {\bf 2016}, \emph{117}, 020402.
\bibitem{dist4} B. Regula, K. Fang, X. Wang, G. Adesso, \PRL {\bf 2018}, \emph{121}, 010401.
\bibitem{dist5} K. Fang, X. Wang, L. Lami, B. Regula, G. Adesso, \PRL {\bf 2018}, \emph{121}, 070404.
\bibitem{dist6} C. L. Liu, D. L. Zhou, \PRA {\bf 2020}, \emph{101}, 012313.
\bibitem{rcc} T. Ma, M. J. Zhao, S. M. Fei, G. L. Long, \PRA {\bf 2016}, \emph{94}, 042312.

%31-40
\bibitem{mc1} Y. Yao, G. H. Dong, L. Ge, M. Li, C. P. Sun, \PRA {\bf 2016}, \emph{94}, 062339.
\bibitem{mc2} M. L. Hu, S. Q. Shen, H. Fan, \PRA {\bf 2017}, \emph{96}, 052309.
\bibitem{mc3} A. Streltsov, H. Kampermann, S. W\"{o}lk, M. Gessner, D. Bru{\ss}, \NJP {\bf 2018}, \emph{20}, 053058.
\bibitem{mirror} S. Kuhr, S. Gleyzes, C. Guerlin, J. Bernu, U. B. Hoff, S. Del\'{e}glise, S. Osnaghi, M. Brune, J.-M. Raimondb, \APL {\bf 2007}, \emph{90}, 164101.
\bibitem{Maniscalco} S. Maniscalco, F. Francica, R. L. Zaffino, N. L. Gullo, F. Plastina, \PRL {\bf 2008}, \emph{100}, 090503.
\bibitem{manzx} Z. X. Man, N. B. An, Y. J. Xia, \PRA {\bf 2014}, \emph{90}, 062104.
\bibitem{Breuer} H.-P. Breuer, F. Petruccione, \textit{The Theory of Open Quantum Systems}, Oxford University Press, Oxford, New York {\bf 2002}.
\bibitem{BLP} H. P. Breuer, E.-M. Laine, J. Piilo, \PRL {\bf 2009}, \emph{103}, 210401.
\bibitem{Bellomo} B. Bellomo, R. L. Franco, G. Compagno, \PRL {\bf 2007}, \emph{99}, 160502.
\bibitem{entan1} B. Bellomo, R. L. Franco, S. Maniscalco, G. Compagno, \PRA {\bf 2008}, \emph{78}, 060302.

%41-50
\bibitem{entan2} B. Bellomo, R. L. Franco, G. Compagno, \PRA {\bf 2008}, \emph{78}, 062309.
\bibitem{loren1} B. Wang, Z. Y. Xu, Z. Q. Chen, M. Feng, \PRA {\bf 2010}, \emph{81}, 014101.
\bibitem{loren2} F. F. Fanchini, T. Werlang, C. A. Brasil, L. G. E. Arruda, A. O. Caldeira, \PRA {\bf 2010}, \emph{81}, 052107.
\bibitem{loren3} L. Mazzola, S. Maniscalco, J. Piilo, K.-A. Suominen, B. M. Garraway, \PRA {\bf 2009}, \emph{79}, 042302.
\bibitem{eur} M. L. Hu, H. Fan, \PRA {\bf 2012}, \emph{86}, 032338.
\bibitem{Zhangyj} Y. J. Zhang, W. Han, Y. J. Xia, Y. M. Yu, H. Fan, \SR {\bf 2015}, \emph{5}, 13359.
\bibitem{deco} M. L. Hu, H. Fan, \SCPMA {\bf 2020}, \emph{63}, 230322.
\bibitem{cpb} M. L. Hu, Y. H. Zhang, H. Fan, \CPB {\bf 2021}, \emph{30}, 030308.
\bibitem{feedback} A. R. R. Carvalho, J. J. Hope, \PRA {\bf 2007}, \emph{76}, 010301.
\bibitem{fro1} T. R. Bromley, M. Cianciaruso, G. Adesso, \PRL {\bf 2015}, \emph{114}, 210401.

%51-60
\bibitem{fro2} M. L. Hu, H. Fan, \SR {\bf 2016}, \emph{6}, 29260.
\bibitem{fro3} X. D. Yu, D. J. Zhang, C. L. Liu, D. M. Tong, \PRA {\bf 2016}, \emph{93}, 060303(R).
\bibitem{new1} J. Yin, Y. H. Li, S. K. Liao, M. Yang, Y. Cao, L. Zhang, J. G. Ren, W. Q. Cai, W. Y. Liu, S. L. Li, R. Shu, Y. M. Huang, L. Deng, L. Li, Q. Zhang, N. L. Liu, Y. A. Chen, C. Y. Lu, X. B. Wang, F. Xu, J. Y. Wang, C. Z. Peng, A. K. Ekert, J. W. Pan, \emph{Nature} {\bf 2020}, \emph{582}, 501.
\bibitem{new2} Y. F. Jiao, S. D. Zhang, Y. L. Zhang, A. Miranowicz, L. M. Kuang, H. Jing, \PRL {\bf 2020}, \emph{125}, 143605.
\bibitem{new3} I. Bloch, \emph{Nature} {\bf 2008}, \emph{453}, 1016.
\bibitem{new4} W. Qin, A. Miranowicz, H. Jing, F. Nori, \PRL {\bf 2021}, \emph{127}, 093602.
\bibitem{new5} W. C. Wang, Y. L. Zhou, H. L. Zhang, J. Zhang, M. C. Zhang, Y. Xie, C. W. Wu, T. Chen, B. Q. Ou, W. Wu, H. Jing, P. X. Chen, \PRA {\bf 2021}, \emph{103}, L020201.
\bibitem{new6} Y. Dai, Y. Dong, Z. Xu, W. You, C. Zhang, O. G\"{u}hne, \emph{Phys. Rev. Appl.} {\bf 2020}, \emph{13}, 054022.
\bibitem{new7} Y. T. Wang, J. S. Tang, Z. Y. Wei, S. Yu, Z. J. Ke, X. Y. Xu, C. F. Li, G. C. Guo, \PRL {\bf 2017}, \emph{118}, 020403.
\bibitem{new8} F. Ming, D. Wang, L. J. Li, X. G. Fan, X. K. Song, L. Ye, J. L. Chen, \AQT {\bf 2021}, \emph{4}, 2100036.

%60-66
\bibitem{EoF1} S. Hill, W. K. Wootters, \PRL {\bf 1997}, \emph{78}, 5022.
\bibitem{EoF2} W. K. Wootters, \PRL {\bf 1998}, \emph{80}, 2245.
\bibitem{new9} F. Ming, W. N. Shi, X. G. Fan, L. Ye, D. Wang, \JPA {\bf 2021}, \emph{54}, 215302.
\bibitem{ohmic} A. J. Leggett, S. Chakravarty, A. T. Dorsey, M. P. A. Fisher, A. Garg, W. Zwerger, \RMP {\bf 1987}, \emph{59}, 1.
\bibitem{finiteT1} S. Shresta, C. Anastopoulos, A. Dragulescu, B. L. Hu, \PRA {\bf 2005}, \emph{71}, 022109.
\bibitem{finiteT2} B. Bellomo, R. L. Franco, G. Compagno, \PRA {\bf 2008}, \emph{77}, 032342.


\end{thebibliography}
%\Non-BibTeX users please use

\end{document}